\documentclass[12pt]{article}
\usepackage{amsmath}
\usepackage{amssymb}
\usepackage{amsfonts}

\renewcommand{\theequation}{\arabic{section}.\arabic{equation}}

\title{Connection between quantum systems involving the fourth Painlev\'e transcendent and $k$-step rational extensions of the harmonic oscillator related to Hermite EOP}

\author{Ian Marquette$^{1,}$\thanks{Electronic address: i.marquette@uq.edu.au} \ and Christiane Quesne$^{2,}$\thanks{Electronic address: cquesne@ulb.ac.be}\\
{\small\sl $^1$ School of Mathematics and Physics, The University of Queensland,}\\
{\small \sl Brisbane, QLD 4072, Australia}\\
{\small\sl $^2$ Physique Nucl\'eaire Th\'eorique et Physique Math\'ematique, 
Universit\'e Libre de Bruxelles,} \\ 
{\small \sl Campus de la Plaine CP229, Boulevard~du Triomphe, B-1050
Brussels, Belgium}}
\date{ }
\begin{document}
\baselineskip=22pt plus 1pt minus 1pt
\maketitle

\begin{abstract}
The purpose of this communication is to point out the connection between a 1D quantum Hamiltonian involving the fourth Painlev\'e transcendent P$_{\rm IV}$, obtained in the context of second-order supersymmetric quantum mechanics and third-order ladder operators, with a hierarchy of families of quantum systems called $k$-step rational extensions of the harmonic oscillator and related with multi-indexed $X_{m_{1},m_{2},...,m_{k}}$ Hermite exceptionnal orthogonal polynomials of type III. The connection between these exactly solvable models is established at the level of the equivalence of the Hamiltonians using rational solutions of the fourth Painlev\'e equation in terms of generalized Hermite and Okamoto polynomials. We also relate the different ladder operators obtained by various combinations of supersymmetric constructions involving Darboux-Crum and Krein-Adler supercharges, their zero modes and the corresponding energies. These results will demonstrate and clarify the relation observed for a particular case in previous papers.

\end{abstract}

\vspace{0.5cm}


 
\newpage
%
%
\section{INTRODUCTION}

The six Painlev\'e transcendents associated with their six corresponding Painlev\'e equations were obtained by Painlev\'e and Gambier \cite{inc56} at the beginning of the 20th century and since then have been the object of many works. They possess many properties, in particular they have no movable branch point \cite{gro02}. They appear in the study and classification of ordinary differential equations of the form $w''=F(z,w,w')$, where $F$ is rational in $w'$ and $w$ and analytic in the variable $z$. Generalizations to higher-order equations have been discussed by Chazy \cite{chazy11}, Bureau \cite{bur64}, and Cosgrove \cite{cos06,cos00}, but the classification and study is a much more difficult problem. \par
%
%
The six Painlev\'e transcendents also appear in various contexts as reductions by symmetry of equations such as KdV, Boussinesq, and soliton, in the fields of statistical mechanics, general relativity, quantum field theory, and optics, for instance \cite{Abl91}. The Painlev\'e transcendents have attracted a lot of interest from both applied and pure mathematics points of view. Their numerous properties, such as their asymptotic, solutions, and irreducibility, have been the topic of several works. \par
%
%
One of these highly interesting objects is the fourth Painlev\'e transcendent $w$ satisfying the fourth Painlev\'e equation
\begin{equation}
  w''=\frac{w'^{2}}{2w}+\frac{3}{2}w^{3}+4 z w^{2} +2 ( z^{2}-\alpha)w+ \frac{\beta}{w}, \label{eq:p4}
\end{equation}
where $\alpha$ and $\beta$ are complex parameters. The fourth Painlev\'e transcendent has been shown to be irreducible, i.e., the general solution of the fourth Painlev\'e equation cannot be expressed in terms of known elementary functions. However, it is known that this fourth Painlev\'e equation possesses hierarchies of rational solutions for special values of the parameters $\alpha$ and $\beta$. It has been shown there are three families of rational solutions of the form
\begin{equation} 
\begin{split}
  w_{1}(z;\alpha_{1},\beta_{1}) & =\frac{P_{1,n-1}(z)}{Q_{1,n}(z)}, \\
  w_{2}(z;\alpha_{2},\beta_{2}) & = -2z+\frac{P_{2,n-1}(z)}{Q_{2,n}(z)},\\
  w_{3}(z;\alpha_{3},\beta_{3}) & = -\frac{2}{3}z+\frac{P_{3,n-1}(z)}{Q_{3,n}(z)},
\end{split}
\end{equation}
where the polynomials $P_{i,n}$ and $Q_{i,n}$ are of degree $n$. The associated values of the parameters $\alpha_{i}$ and $\beta_{i}$ ($i=1$, 2, 3) are summarized in Table I. These three hierarchies of rational solutions of the fourth Painlev\'e equation are known as the ``$-\frac{1}{z}$'', ``$-2z$'', and ``$-\frac{2}{3}z$'' hierarchies.\par
%
%
\begin{table}[h!]

\caption{Three families of rational solutions and corresponding parameters.}

\begin{center}
\begin{tabular}{llllll}
  \hline\hline\\[-0.2cm]
   Family type & Hierarchy & $\alpha_{i}$ & $\beta_{i}$ & $m$ & $n$\\[0.2cm]\hline\\[-0.2cm]
  1 &``$-\frac{1}{z}$'' & $\pm m$ & $-2(1+2n +m)^{2}$ & $m \in \mathbb{Z}$, $m \geq -2n$  & $n \in 
        \mathbb{Z}$, $n \leq -1$ \\[0.2cm]
  2&``$-2z$'' & $m$ & $-2(1+2n +m)^{2}$ & $m \in \mathbb{Z}$, $m \geq -n$ & $n \in \mathbb{Z}$, 
        $n \geq 0$\\[0.2cm]
  3&``$-\frac{2}{3}z$'' & $m$  & $\frac{2}{9}(1+ 6n -3m)^{2}$ & $m \in \mathbb{Z}$ & $n \in \mathbb{Z}$ 
        \\[0.2cm]
  \hline\hline 
\end{tabular}
\end{center}

\end{table}
\par
%
%
Very interestingly, the fourth Painlev\'e transcendent appears in the context of nonrelativistic quantum systems and, more particularly, of second-order supersymmetric quantum mechanics and third-order ladder operators  \cite{andri00, nik97, ves93, ves01, carb04, mat08, ber11a}. Andrianov, Cannata, Ioffe, and Nishnianidze \cite{andri00} have considered a set of two Hamiltonians $H_1$, $H_2$, which are related at the same time by first- and second-order Darboux transformations and can be expressed as
\begin{equation}
  H_{1,2}=-\frac{d^{2}}{dx^{2}}+ x^{2} \mp g' + g^{2} + 2 x g -1,  \label{eq:h12p4}
\end{equation} 
where $g=w(x,\alpha,\beta)$ (with $-\infty < x < \infty$) satisfies the fourth Painlev\'e equation (\ref{eq:p4}) for arbitrary parameters $\alpha$ and $\beta$. Note that, in Ref.~\cite{andri00}, the parameters 
\begin{equation}
  a=\alpha, \qquad b=-2\beta, \qquad \bar{\alpha} = a-1 = \alpha-1, \qquad d=-\frac{1}{4}b = \frac{1}{2}
  \beta  \label{eq:para-andri}
\end{equation}
are actually used and that we set there $\lambda=1$ for simplicity's sake. The Hamiltonians $H_{1}$ and $H_{2}$ admit several symmetries, which allow us to restrict the discussion to Hamiltonian $H_{1}$ only. It has been shown by the authors of \cite{andri00} that for some ranges of $\alpha$ and $\beta$ values, $H_1$ may admit one, two, or three infinite sequences of equidistant levels, or one infinite sequence of equidistant levels with either one additional singlet or one additional doublet. This system is still the object of works in regard of specific choices of parameters, non-Hermitian Hamiltonians, ladder operators, and coherent states \cite{ber11b, ber15, fer15}. Superintegrable generalizations of this model in two dimensions, which possess higher-order integrals of motion and finitely generated polynomial algebras, have also been obtained and studied algebraically \cite{gra04,marquette09a,marquette09b,marquette09c,marquette10}.\par
%
%
More recently, multi-step rational extensions of the harmonic oscillator were studied \cite{fel09, oda11, oda13, ull14, ull13, marquette13a, marquette13b, marquette13c, marquette14}. The corresponding Hamiltonians may be written as \cite{marquette14, footnote}
\begin{equation}
  H^{(2)} =-\frac{d^{2}}{dx^{2}}+  x^2 - 2k - 2 \frac{d^2}{dx^2} \log {\cal W}({\cal H}_{m_1}, {\cal H}_{m_2},
  \ldots,  {\cal H}_{m_k}),   \label{eq:hkstep}
\end{equation}
in terms of Wronskians of pseudo-Hermite polynomials ${\cal H}_{m_i}(x)$ \cite{fel09}. Equation (\ref{eq:hkstep}) corresponds in fact to a hierarchy of families indexed by the parameter $k$, i.e., $k=1$, 2, 3,\ldots. A given family of models corresponding to a given value $k=l$ with $l \in \mathbb{Z}$  is a $l$-step extension that contains $l$ parameters $m_{1}, \ldots, m_{l}$ satisfying some conditions. The wavefunctions of these models are related to multi-indexed $X_{m_{1},m_{2},...,m_{l}}$ Hermite exceptional orthogonal polynomials (EOP) of type III. This hierarchy can be obtained via various supersymmetric constructions according to the Darboux-Crum \cite{crum} or Krein-Adler \cite{krein, adler} approach. These different constructions also make possible to generate different sets of ladder operators and, consequently, different sets of integrals of motion for their superintegrable generalizations in 2D \cite{marquette13a, marquette13b, marquette13c, marquette14}.\par
%
%
The purpose of this paper is to relate both Hamiltonians, i.e., $H^{(2)}$ for specific values of $k$ and $H_{1}$  for hierarchies of rational solutions of the fourth Painlev\'e transcendent, and in addtion to relate the supercharges involved in their distinct supersymmetric schemes and the corresponding ladder operators associated with some combinations of supercharges and different structures of zero modes. The aim of this communication is also to fill in a gap in the literature and explicitly demonstrate the connection that was observed and mentioned in recent papers in regard of a particular case of $H_{1}$ and the one-step rational extension $H^{(2)}$ for $m_1=2$ \cite{marquette09a, marquette09b, marquette13a, marquette13b, post15}. \par
%
%
The organization of the paper is as follows. In Sec.~II, we recall some results on $k$-step rational extensions of the harmonic oscillator $H^{(2)}$ and consider more explicitly one- and two-step rational extensions.  In Sec.~III, we discuss the quantum systems $H_{1,2}$ involving the fourth Painlev\'e transcendent, the construction of ladder operators of Andrianov et al.\ for $H_{1}$ and recall some known results on the three hierarchies of rational solutions related to generalized Hermite and Okamoto polynomials. In Sec.~IV, we discuss for $H_{1}$ the case of one singlet and one infinite chain and show the connection with one-step rational extensions $H^{(2)}$. We compare the supercharges, the ladder operators, the zero modes and the corresponding energy spectra. In Sec.~V, we connect the case of three infinite sequences for $H_{1}$ with the particular case $m_{1}=2$ for one-step rational extensions via generalized Okamoto polynomials. In Sec.~VI,  we relate the case of one doublet and one infinite sequence of levels for $H_{1}$ to the two-step rational extension case for $H^{(2)}$ and specific values of parameters. In Sec.~VII, we present some concluding remarks.\par
%
%
\section{\boldmath $k$-STEP RATIONAL EXTENSIONS OF THE HARMONIC OSCILLATOR}
\setcounter{equation}{0}

In $n$th-order supersymmetric quantum mechanics \cite{fernandez04}, one considers a pair of partner Hamiltonians $H^{(i)} = - \frac{d^2}{dx^2} + V^{(i)}(x)$, $i=1$, 2, where $V^{(2)}(x)$ can be obtained from $V^{(1)}(x)$ in the form
\begin{equation}
  V^{(2)}(x) = V^{(1)}(x) - 2 \frac{d^2}{dx^2} \log {\cal W}(\varphi_1, \varphi_2, \ldots, \varphi_n),
  \label{eq:partner}
\end{equation}
by considering $n$ different seed solutions $\varphi_1(x)$, $\varphi_2(x)$, \ldots, $\varphi_n(x)$ of the Schr\"odinger equation for $H^{(1)}$. The two Hamiltonians $H^{(1)}$ and $H^{(2)}$ intertwine with two $n$th-order differential operators $\cal A$ and ${\cal A}^{\dagger}$ as 
\begin{equation}
  {\cal A} H^{(1)} = H^{(2)} {\cal A}, \qquad {\cal A}^{\dagger} H^{(2)} = H^{(1)} {\cal A}^{\dagger}.
\end{equation}
Here ${\cal A} = A^{(n)} \cdots A^{(2)} A^{(1)}$, ${\cal A}^{\dagger} = A^{(1)\dagger} A^{(2)\dagger} A^{(n)\dagger}$, and
\begin{equation}
\begin{split}
  & A^{(i)} = \frac{d}{dx} + W^{(i)}(x), \qquad W^{(i)}(x) = - \frac{d}{dx} \log \varphi^{(i)}(x), \qquad i=1, 2, 
       \ldots, n, \\
  & \varphi^{(1)}(x) = \varphi_1(x), \qquad \varphi^{(i)}(x) = \frac{{\cal W}(\varphi_1, \varphi_2, \ldots, 
       \varphi_i)}{{\cal W}(\varphi_1, \varphi_2, \ldots, \varphi_{i-1})}, \qquad i=2, 3, \ldots, n. 
\end{split}
\end{equation}
In the present case, the starting Hamiltonian is that of the harmonic oscillator, i.e., $V^{(1)}(x) = x^2$, $-\infty < x < \infty$.\par
%
%
In the state-adding case (or Darboux-Crum approach \cite{crum}), we choose $n=k$ and take for $\varphi_i(x)$ nonnormalizable  eigenfunctions $\phi_{m_i}(x)$ of $H^{(1)}$ below the ground state, where 
\begin{equation}
  \phi_m(x) = {\cal H}_m(x) e^{\frac{1}{2} x^2}  \label{eq:phi}
\end{equation}
is expressed in terms of a $m$th-degree pseudo-Hermite polynomial ${\cal H}_m(x) = (-{\rm i})^m H_m({\rm i}x)$ \cite{fel09}. As a result, we get the potential of a $k$-step rational extension of the harmonic oscillator
\begin{equation}
  V^{(2)}(x) = x^2 - 2k - 2 \frac{d^2}{dx^2} \log {\cal W}({\cal H}_{m_1}, {\cal H}_{m_2}, \ldots, 
  {\cal H}_{m_k}),  \label{eq:vkstep}
\end{equation}
corresponding to Hamiltonian (\ref{eq:hkstep}). Non-singularity of $V^{(2)}(x)$ can be achieved by taking $m_1 < m_2 < \cdots < m_k$ with $m_i$ even (resp.\ odd) for $i$ odd (resp.\ even). Hamiltonian $H^{(2)}$ is exactly solvable with wavefunctions related to Hermite EOP of type III. The intertwining operators (also called supercharges) $\cal A$ and ${\cal A}^{\dagger}$ are then $k$th-order differential operators. As explicitly shown in the $k=2$ case \cite{marquette13b}, other orderings of $m_1$, $m_2$, \ldots, $m_k$ are possible and lead to other supercharge operators.\par
%
%
Up to some additive constant, the same potential $V^{(2)}(x)$ can be constructed \cite{marquette14} via the state-deleting or Krein-Adler approach \cite{krein, adler}. In such a case, we choose $n = m_k +1 -k$ and take for $\varphi_i(x)$ some well-selected bound-state wavefunctions of $H^{(1)}$. This results in another set of supercharges, which are ($m_k+1-k$)th-order differential operators.\par
%
%
As it will be detailed below, ladder operators for $H^{(2)}$ can be constructed by appropriately combining supercharges.\par
%
%
Let us now present some explicit expressions for the potentials, spectra, wavefunctions, supercharges, and ladder operators corresponding to the one- and two-step cases, which we plan to consider in the present paper.\par
%
%
\subsection{One-step rational extensions}

{}For $k=1$, the potential $V^{(2)}(x)$ of Eq.~(\ref{eq:vkstep}), obtained in the state-adding approach, takes the form \cite{marquette13a, marquette13b, marquette13c}
\begin{equation}
  V^{(2)}(x) = x^2 - 2 \left[\frac{{\cal H}''_{m_1}}{{\cal H}_{m_1}} - \left(\frac{{\cal H}'_{m_1}}
  {{\cal H}_{m_1}}\right)^2 + 1 \right], \label{eq:v1step}
\end{equation}
where $m_1$ is assumed even for non-singularity's sake. The corresponding supercharges $\cal A$ and ${\cal A}^{\dagger}$ reduce to 
\begin{equation}
  A=\frac{d}{dx} + W(x), \qquad A^{\dagger}=-\frac{d}{dx}+W(x), \qquad W(x)=-x-\frac{\mathcal{H}'_{m_1}}
  {\mathcal{H}_{m_1}},  \label{eq:AAdW}
\end{equation}
and the Hamiltonian $H^{(2)}$ can be written as 
\begin{equation}
  H^{(2)} = A A^{\dagger} - 2m_1 - 1 \qquad \text{with} \qquad V^{(2)} = W^2 + W' - 2m_1 - 1.  
  \label{eq:haa}
\end{equation}
\par
%
%
The spectrum of $H^{(2)}$ is now given by 
\begin{equation}
  E^{(2)}_{\nu} = 2\nu+1, \qquad \nu=-m_1-1, 0, 1, 2, \ldots,  \label{eq:energ1s}
\end{equation}
and differs from that of the standard harmonic oscillator $V^{(1)}(x)$ by an extra level below the infinite sequence of levels. The corresponding wavefunctions are expressed as
\begin{equation}
  \psi^{(2)}_{\nu}(x) \propto \frac{e^{-\frac{1}{2}x^{2}}}{\mathcal{H}_{m_1}(x)} y_{n}^{(m_1)}(x), \qquad 
  n=m_1+\nu+1, \label{eq:wave1s}
\end{equation}
where the $n$th-degree polynomials $y_{n}^{(m_1)}(x)$, given by
\begin{equation} 
  y^{m_1}_n(x) = \begin{cases}
    1 & \text{if $\nu = -m_1-1$}, \\
    -\mathcal{H}_{m_1}H_{\nu+1}-2m_1 \mathcal{H}_{m_1-1}H_{\nu} & \text{if $\nu=0$, 1, 2, \ldots,} 
    \end{cases}  \label{eq:y1step}
\end{equation}
belong to the Hermite EOP family of type III that can be denoted as $X_{m_1}$. \par
%
%
On the other hand, by deleting the first $m_1$ excited states of $H^{(1)}$, $\psi_1(x)$, $\psi_2(x)$, \ldots, $\psi_{m_1}(x)$, where $\psi_i(x) \propto e^{-x^2/2} H_i(x)$, we get the potential $\bar{V}^{(2)}(x) = V^{(2)}(x) + 2m_1 + 2$ and another set of supercharges $\bar{\cal{A}} = \bar{A}^{(m_1)} \cdots \bar{A}^{(2)} \bar{A}^{(1)}$ and $\bar{\cal{A}}^{\dagger} = \bar{A}^{(1)\dagger} \bar{A}^{(2)\dagger} \cdots \bar{A}^{(m_1)\dagger}$ \cite{marquette14}, where
\begin{equation}
  \bar{A}^{(i)} = \frac{d}{dx} + \bar{W}^{(i)}(x), \qquad \bar{W}^{(i)}(x) = x + \frac{{\cal H}'_{i-1}}
  {{\cal H}_{i-1}} - \frac{{\cal H}'_i}{{\cal H}_i}, \quad i=1, 2, \ldots, m_1.  \label{eq:Abar}
\end{equation}
\par
%
%
Two different classes of ladder operators are known for one-step rational extensions of the harmonic oscillator. The first ones, corresponding to the standard way of building ladder operators in supersymmetric quantum mechanics, are obtained by combining the oscillator creation and annihilation operators $a^{\dagger} = - d/dx + x$, $a = d/dx + x$ with the supercharges $A$ and $A^{\dagger}$ of Eq.~(\ref{eq:AAdW}). They are given by \cite{marquette13a, marquette13c}
\begin{equation}
  b^{\dagger} = A a^{\dagger} A^{\dagger}, \qquad b = A a A^{\dagger},  \label{eq:b}
\end{equation}
and satisfy the relations $[H^{(2)}, b^{\dagger}] = 2 b^{\dagger}$, $[H^{(2)}, b] = - 2 b$. They are third-order operators for any (even) $m_1$ value. The annihilation operator $b$ has two zero modes $\psi^{(2)}_{-m_1-1}(x)$ and $\psi^{(2)}_0(x)$, while the creation operator $b^{\dagger}$ has one zero mode $\psi^{(2)}_{-m_1-1}(x)$, so that the set of wavefunctions divides into a singlet $\psi^{(2)}_{-m_1-1}(x)$ and an infinite sequence $\{\psi^{(2)}_{\nu}(x) \mid \nu=0,1,2,\ldots\}$, corresponding to equidistant levels.\par
%
%
The second class of ladder operators is obtained by combining the supercharges (\ref{eq:AAdW}) and (\ref{eq:Abar}) of the state-adding and state-deleting approaches. They are $(m_1+1)$th-order operators given by \cite{marquette13c, marquette14}
\begin{equation}
  c^{\dagger} = A \bar{A}^{(1)\dagger} \bar{A}^{(2)\dagger} \cdots \bar{A}^{(m_1)\dagger}, \qquad
  c = \bar{A}^{(m_1)} \cdots \bar{A}^{(2)} \bar{A}^{(1)} A^{\dagger},  \label{eq:c}
\end{equation}
and satisfying the relations $[H^{(2)}, c^{\dagger}] = (2m_1+2) c^{\dagger}$, $[H^{(2)}, c] = - (2m_1+2) c$. In Sec.~V, we will consider more specifically the $m_1=2$ case, wherein the ladder operators $c^{\dagger}$, $c$ are third-order operators. The annihilation operator $c$ has then three zero modes $\psi^{(2)}_{-3}(x)$, $\psi^{(2)}_1(x)$, $\psi^{(2)}_2(x)$, while the creation operator $c^{\dagger}$ has none. Hence, the set of wavefunctions divides into three infinite subsets $\{\psi^{(2)}_{i+3 j}(x) \mid j=0,1,2,\ldots\}$ with $i=-3$, 1, 2, corresponding to equidistant levels.\par
%
%
\subsection{Two-step rational extensions}

{}For $k=2$, the potential $V^{(2)}(x)$, obtained in the state-adding case, reads \cite{marquette13b, marquette14}
\begin{equation}
  V^{(2)}(x) = x^2 - 2 \left[\frac{g''_{\mu}}{g_{\mu}} - \left(\frac{g'_{\mu}}{g_{\mu}}\right)^2 + 2 \right],
  \label{eq:v2step}
\end{equation}
where $g_{\mu}(x) \equiv {\cal W}({\cal H}_{m_1}, {\cal H}_{m_2})$ is a $\mu$th-degree polynomial (with $\mu = m_1+m_2-1$) and we assume that $m_1$ is even, $m_2$ is odd, and $m_1 < m_2$. The spectrum of $H^{(2)}$ is given by
\begin{equation}
  E^{(2)}_{\nu} = 2\nu + 1, \qquad \nu=-m_2-1, -m_1-1, 0, 1, 2, \ldots, \label{eq:energ2s}
\end{equation}
so that there are two extra levels below the standard oscillator spectrum, the corresponding wavefunctions being
\begin{equation}
  \psi^{(2)}_{\nu}(x) \propto \frac{e^{-\frac{1}{2}x^2}}{g_{\mu}(x)} y_n^{(\mu)}(x), \qquad n=\mu+\nu+2. 
  \label{eq:wave2s}
\end{equation}
The $n$th-degree polynomials $y^{(\mu)}_n(x)$, which read
\begin{equation} 
  y^{(\mu)}_n(x) = \begin{cases}
     \mathcal{H}_{m_1}, & \text{if $\nu= -m_2-1$}, \\
     \mathcal{H}_{m_2}, & \text{if $\nu= -m_1-1$}, \\
     (m_2-m_1) \mathcal{H}_{m_1}\mathcal{H}_{m_2}H_{\nu+1} & \\
     \quad {} + 2[m_1(m_2+\nu+1)\mathcal{H}_{m_1-1}
         \mathcal{H}_{m_2} & \\
     \quad {} - m_2(m_1+\nu+1)\mathcal{H}_{m_1}\mathcal{H}_{m_2-1}]H_{\nu},
         & \text{if $\nu=0$, 1, 2, \ldots},         
  \end{cases}
\end{equation}
belong to the Hermite EOP of type III denoted as $X_{m_1,m_2}$. \par
%
%
We will not discuss here the state-deleting approach because the combination of the corresponding supercharges with those coming from the state-deleting approach always leads to ladder operators of order higher than three, but we will instead consider two different ways of adding the states with $\nu=-m_1-1$ and $\nu=-m_2-1$ below the standard oscillator spectrum. Adding first the state with $\nu=-m_1-1$ leads to an intermediate Hamiltonian with a non-singular potential and supercharge operators ${\cal A} = A^{(2)} A^{(1)}$, ${\cal A}^{\dagger} = A^{(1)\dagger} A^{(2)\dagger}$ with
\begin{equation}
  A^{(i)} = \frac{d}{dx} + W^{(i)}(x), \qquad W^{(1)}(x) = - x - \frac{{\cal H}'_{m_1}}{{\cal H}_{m_1}}, \qquad
  W^{(2)}(x) = - x + \frac{{\cal H}'_{m_1}}{{\cal H}_{m_1}} - \frac{g'_{\mu}}{g_{\mu}}.
\end{equation}
Starting instead by adding the state with $\nu=-m_2-1$ yields an intermediate Hamiltonian with a singular potential (the final Hamiltonian being the same as before) and other supercharge operators $\tilde{{\cal A}} = \tilde{A}^{(2)} \tilde{A}^{(1)}$, $\tilde{{\cal A}}^{\dagger} = \tilde{A}^{(1)\dagger} \tilde{A}^{(2)\dagger}$ with
\begin{equation}
  \tilde{A}^{(i)} = \frac{d}{dx} + \tilde{W}^{(i)}(x), \qquad \tilde{W}^{(1)}(x) = - x 
  - \frac{{\cal H}'_{m_2}}{{\cal H}_{m_2}}, \qquad \tilde{W}^{(2)}(x) = - x 
  + \frac{{\cal H}'_{m_2}}{{\cal H}_{m_2}} - \frac{g'_{\mu}}{g_{\mu}}.
\end{equation}
As shown in \cite{marquette13b}, the two intermediate Hamiltonians intertwine with the $(m_2-m_1)$th-order operators $\hat{A}_{m_2-m_1} \cdots \hat{A}_2 \hat{A}_1$ and $\hat{A}^{\dagger}_1 \hat{A}^{\dagger}_2 \cdots \hat{A}^{\dagger}_{m_2-m_1}$, where 
\begin{equation}
  \hat{A}_i = \frac{d}{dx} + \hat{W}_i(x), \qquad \hat{W}_i(x) = x + \frac{{\cal H}'_{m_1+i-1}}
  {{\cal H}_{m_1+i-1}} - \frac{{\cal H}'_{m_1+i}}{{\cal H}_{m_1+i}}, \qquad i=1, 2, \ldots, m_2-m_1.
\end{equation}
Combining the supercharges gives rise to $(m_2-m_1+2)$th-order ladder operators
\begin{equation}
  d^{\dagger} = A^{(2)} \hat{A}^{\dagger}_1 \hat{A}^{\dagger}_2 \cdots \hat{A}^{\dagger}_{m_2-m_1}
  \tilde{A}^{(2)\dagger}, \qquad d = \tilde{A}^{(2)} \hat{A}_{m_2-m_1} \cdots \hat{A}_2 \hat{A}_1
  A^{(2)\dagger},  \label{eq:dpd}  
\end{equation}
satisfying the relations $[H^{(2)}, d^{\dagger}] = 2(m_2-m_1) d^{\dagger}$, $[H^{(2)}, d] = - 2(m_2-m_1) d$. In Sec.~VI, we will restrict ourselves to the $m_2 = m_1+1$ case, wherein the ladder operators $d^{\dagger}$, $d$ are third-order operators. The annihilation operator $d$ has then two zero modes $\psi^{(2)}_{-m_1-2}(x)$, $\psi^{(2)}_0(x)$, while the creation operator $d^{\dagger}$ has a single zero mode $\psi^{(2)}_{-m_1-1}(x)$, so that the set of wavefunctions divides into a doublet $\{\psi^{(2)}_{-m_1-2}(x), \psi^{(2)}_{-m_1-1}(x)\}$ and an infinite sequence $\{\psi^{(2)}_{\nu}(x) \mid \nu=0,1,2,\ldots\}$, corresponding to equidistant levels.\par
%
%
\section{\boldmath HAMILTONIANS $H_{1,2}$ WITH FOURTH PAINLEVE TRANSCENDENT}
\setcounter{equation}{0}

The two Hamiltonians $H_1$ and $H_2$, given by Eq.~(\ref{eq:h12p4}), are related by the following first-order supercharges \cite{andri00}
\begin{equation}
  q^+ = \frac{d}{dx} + W_3, \qquad q^- = - \frac{d}{dx} + W_3,
\end{equation}
defined in terms of some superpotential $W_3(x)$, and they are assumed to satisfy the intertwining relations
\begin{equation}
  H_1 q^+ = q^+ (H_2+2), \qquad q^- H_1 = (H_2+2) q^-.  \label{eq:intertwine1}
\end{equation}
As a consequence, we can write $H_1 = q^+ q^-$, $H_2 = q^- q^+ - 2$, and in particular
\begin{equation}
  H_1 = - \frac{d^2}{dx^2} + W_3^2 + W_3'.  \label{eq:h1}
\end{equation}
There is a second set of second-order supercharges, denoted by $M^+$ and $M^-$, and satisfying the intertwining relations
\begin{equation}
  H_1 M^+ = M^+ H_2, \qquad M^- H_1 = H_2 M^-,  \label{eq:intertwine2}
\end{equation}
so that the systems $H_1$ and $H_2$ are related by two supersymmetric quantum mechanical structures, one with first-order supercharges and another one with second-order supercharges.\par
%
%
{}For the second-order supercharges $M^{\pm}$, one may distinguish two cases: the reducible case, in which the supercharges factorize into products of first-order operators with real superpotentials $W_1$ and $W_2$, and the irreducible one, in which they do not factorize in this way. The latter corresponds to positive values of parameter $d$, while the former occurs whenever $d \le 0$. We will be concerned here only with the reducible case, wherein
\begin{equation}
  M^+ = \left(\frac{d}{dx} + W_1\right) \left(\frac{d}{dx} + W_2\right), \qquad M^- = \left(-\frac{d}{dx} + 
  W_2\right) \left(-\frac{d}{dx} + W_1\right). 
\end{equation}
\par
%
%
The three superpotentials $W_1$, $W_2$, and $W_3$, involved in the construction described by Andrianov et al.~\cite{andri00}, can be written in terms of the function $g$, satisfying the fourth Painlev\'e equation, and its derivative as
\begin{equation} 
\begin{split}
  W_1 & = - \frac{g}{2} + \frac{g'}{2g} - \frac{c}{2g}, \\
  W_2 & = - \frac{g}{2} - \frac{g'}{2g} + \frac{c}{2g}, \\
  W_3 & = - g - x,   \label{eq:w2}
\end{split}
\end{equation}
where $c^2 = - 4d$ and, as a consequence, $c$ may be either $2\sqrt{-d}$ or $-2\sqrt{-d}$ \cite{andrianov95}.\par
%
%
On combining Eqs.~(\ref{eq:intertwine1}) and (\ref{eq:intertwine2}), it results that the third-order operators
\begin{equation}
\begin{split}
  & a^+ = q^+ M^- = \left(\frac{d}{dx} + W_3\right) \left(- \frac{d}{dx} + W_2\right) \left(- \frac{d}{dx} +
      W_1\right), \\
  & a^- = M^+ q^- = \left(\frac{d}{dx} + W_1\right) \left(\frac{d}{dx} + W_2\right) \left(- \frac{d}{dx} +
      W_3\right),
\end{split}  \label{eq:ap-am}
\end{equation}
fulfil the relations $[H_1, a^+] = 2a^+$, $[H_1, a^-] = - 2a^-$, and therefore behave as ladder operators for the Hamiltonian $H_1$ involving the fourth Painlev\'e transcendent.\par
%
%
The annihilation operator $a^-$ and the creation one $a^+$ can both admit three possible zero modes, denoted by ($\psi^{(0)}_0$, $\psi_+^{(0)}$, $\psi_-^{(0)}$) and ($\psi_1$, $\psi_2$, $\psi_3$), respectively. They can be written as
\begin{equation} 
\begin{split}
  \psi_0^{(0)}(x) & = e^{\int^{x} W_3(x') dx'}, \\
  \psi_+^{(0)}(x) & = [W_2(x)-W_3(x)] e^{-\int^{x} W_2(x') dx'}, \\
  \psi_-^{(0)}(x) & = \{c + [W_2(x)-W_3(x)][W_1(x)+W_2(x)]\} e^{-\int^{x} W_1(x') dx'},   
\end{split}  \label{eq:0pm}
\end{equation}
and
\begin{equation} 
\begin{split}
  \psi_1(x) & = e^{\int^{x} W_1(x') dx'}, \\
  \psi_2(x) & =[W_1(x) + W_2(x)] e^{\int^{x} W_2(x') dx'}, \\
  \psi_{3}(x) & = \{E_+^{(0)} + [W_1(x) + W_2(x)] [W_2(x) - W_3(x)]\} e^{-\int^{x} W_3(x')dx'},   
\end{split}  \label{eq:123}
\end{equation} 
with corresponding energies
\begin{equation} 
  E_0^{(0)} = 0, \qquad E_-^{(0)} = \bar{\alpha} + 2 - \tfrac{1}{2} c, \qquad E_+^{(0)} = \bar{\alpha} + 2
  + \tfrac{1}{2} c, \label{eq:energy0pm}
\end{equation}
and 
\begin{equation} 
  E_1 = \bar{\alpha} - \tfrac{1}{2} c, \qquad E_2 = \bar{\alpha} + \tfrac{1}{2} c, \qquad E_3 = -2. 
  \label{eq:energy123}
\end{equation}
\par
%
%
All the relations considered so far in this section are only formal and in every special case one must discuss the normalizability of wavefunctions, which will depend on the values taken by the parameters. In particular, due to conflicting asymptotics, at most three of the six possible zero modes of $a^-$ and $a^+$, given in Eqs.~(\ref{eq:0pm}) and (\ref{eq:123}), can be square integrable at the same time.\par
%
%
In connection with rational extensions of the harmonic oscillator, we will be interested in rational solutions of the fourth Painlev\'e equation, which we now proceed to briefly review.\par
%
%
\subsection{Rational solutions of the fourth Painlev\'e equation}

There exists three distinct families of rational solutions of the fourth Painlev\'e equation, the different members of which can be generated via B\"acklund transformations \cite{fuk00, kam83, nou99, clar03, clar09}.\par
%
%
Two of these families, namely the so-called ``$-1/z$'' and ``$-2z$'' hierarchies, can be written in terms of generalized Hermite polynomials $H_{m,n}(z)$ as
\begin{equation}
  w^{(\rm I)}_{m,n} = w(z;\alpha_{m,n}^{(\rm I)},\beta_{m,n}^{(\rm I)}) = -\frac{d}{dz} 
  \log\left(\frac{H_{m,n+1}}{H_{m,n}}\right),   \label{eq:1wI}
\end{equation}
\begin{equation}
  w^{(\rm II)}_{m,n} = w(z;\alpha_{m,n}^{(\rm II)},\beta_{m,n}^{(\rm II)}) = \frac{d}{dz} 
  \log\left(\frac{H_{m+1,n}}{H_{m,n}}\right),  \label{eq:1wII}
\end{equation}
with corresponding parameters $\alpha^{(\rm I)}_{m,n}$, $\beta^{(\rm I)}_{m,n}$ and $\alpha^{(\rm II)}_{m,n}$, $\beta^{(\rm II)}_{m,n}$ given in Table II.\par
%
%
\begin{table}[h!]

\caption{Parameters of solutions $w^{(\rm I)}_{m,n}$ and $w^{(\rm II)}_{m,n}$ including both ``$-1/z$'' and ``$-2z$'' hierarchies }

\begin{center}
\begin{tabular}{lllll}
  \hline\hline\\[-0.2cm] 
  $w^{(\rm i)}$   & $\alpha^{(\rm i)}$ & $\beta^{(\rm i)}$ & $m$ & $n$   \\[0.2cm]
  \hline \\[-0.2cm]
  $w^{(\rm I)}_{m,n}$  &  $-(m+2n+1)$  &   $-2m^{2}$      & $m \in \mathbb{Z}$ , $m \geq 0$  & $n \in 
       \mathbb{Z}$ ,  $n \geq 0$ \\[0.2cm]
  $w^{(\rm II)}_{m,n}$ & $2m+n+1$      &   $-2n^{2}$      & $m \in \mathbb{Z}$ , $m \geq 0$ & $n \in 
       \mathbb{Z}$ ,  $n \geq 0$  \\[0.2cm]
  \hline\hline 
\end{tabular}
\end{center}

\end{table}
\par
%
%
The third family, which is the so-called ``$-\frac{2}{3}z$'' hierarchy, can be written in terms of generalized Okamoto polynomials $Q_{m,n}(z)$ as 
\begin{equation}
  w^{(\rm I)}_{m,n} = w(z;\alpha_{m,n}^{(\rm I)},\beta_{m,n}^{(\rm I)}) = - \frac{2}{3}z -\frac{d}{dz} 
  \log\left(\frac{Q_{m,n+1}}{Q_{m,n}}\right),   \label{eq:2wI}
\end{equation}
\begin{equation}
  w^{(\rm II)}_{m,n} = w(z;\alpha_{m,n}^{(\rm II)},\beta_{m,n}^{(\rm II)}) = - \frac{2}{3}z + \frac{d}{dz} 
  \log\left(\frac{Q_{m+1,n}}{Q_{m,n}}\right),  \label{eq:2wII}
\end{equation}
with corresponding parameters $\alpha^{(\rm I)}_{m,n}$, $\beta^{(\rm I)}_{m,n}$ and $\alpha^{(\rm II)}_{m,n}$, $\beta^{(\rm II)}_{m,n}$ given in Table III.\par
%
%
\begin{table}[h!]

\caption{Parameters of solutions $w^{(\rm I)}_{m,n}$ and $w^{(\rm II)}_{m,n}$ of the ``$-\frac{2}{3}z$'' hierarchy}

\begin{center}
\begin{tabular}{lllll}
  \hline\hline\\[-0.2cm]
  $w^{(\rm i)}$   & $\alpha^{(\rm i)}$ & $\beta^{(\rm i)}$ & $m$ & $n$  \\[0.2cm]
  \hline\\[-0.2cm]
  $w^{(\rm I)}_{m,n}$  &  $-2n-m$  &   $-\frac{2}{9}(3m-1)^{2}$      & $m \in \mathbb{Z}$, $m \geq 0$  & 
         $n \in \mathbb{Z}$ ,  $n \geq 0$ \\[0.2cm]
  $w^{(\rm II)}_{m,n}$ & $2m+n$      &   $-\frac{2}{9}(3n-1)^{2}$      & $m \in \mathbb{Z}$,  $m \geq 0$  & 
  $n \in \mathbb{Z}$ ,  $n \geq 0$  \\[0.2cm]
  \hline\hline 
\end{tabular}
\end{center}

\end{table}
\par
%
%
The generalized Hermite and Okamoto polynomials satisfy some recurrence relations, which can be written using Hirota operator, and they also admit various representations in terms of determinants \cite{clar03}. Some special cases are given by $H_{n,0}=H_{0,n}=1$, $H_{n,1}(z) = H_n(z)$, $H_{1,n}(z) = {\cal H}_n(z)$, and $Q_{0,0}=Q_{1,0}=Q_{0,1}=1$, $Q_{1,1}(z) = \sqrt{2} z$, $Q_{2,0}(z) = 2z^2+3$, $Q_{0,2}(z) = 2z^2-3$.\par
%
%
\subsection{Generalized Hermite polynomials}

For future purposes, we need to point out some identities relating generalized Hermite polynomials with Wronskians of Hermite or pseudo-Hermite polynomials.\par
%
%
It is known \cite{clar09} that generalized Hermite polynomials can be expressed as Wronskians of standard Hermite polynomials as
\begin{equation}
  H_{m,n}(z) = {\cal W}(H_m(z), H_{m+1}(z), \ldots, H_{m+n-1}(z)) \equiv {\cal W}\left(\{H_{m+j}(z)\}_{j=0}
  ^{n-1}\right), \quad m, n \ge 1,  \label{eq:gen-her-1}
\end{equation}
with $H_{m,0}(z) = H_{0,n}(z) = 1$, so that they are $(mn)$th-degree polynomials.\par
%
%
On the other hand, it has been shown \cite{oda13} that Wronskians of Hermite and pseudo-Hermite polynomials are connected by the relation
\begin{align}
  & {\cal W}({\cal H}_{n_1}(z), {\cal H}_{n_2}(z), \ldots, {\cal H}_{n_m}(z)) \nonumber \\
  & \quad\propto {\cal W}(H_1(z), H_2(z), \ldots, \check{H}_{n_m-n_{m-1}}(z), \ldots, 
     \check{H}_{n_m-n_1}(z), \ldots, H_{n_m}(z)),    
\end{align}
where, on the right-hand side, $H_{n_m-n_{m-1}}(z)$, \ldots, $H_{n_m-n_2}(z)$, $H_{n_m-n_1}(z)$ are excluded from the Wronskian. On choosing $n_i = n+i-1$, $i=1$, 2, \ldots, $m$, in this equation, we can rewrite the generalized Hermite polynomials of Eq.~(\ref{eq:gen-her-1}) as Wronskians of pseudo-Hermite polynomials,
\begin{equation}
  H_{m,n}(z) = {\cal W}({\cal H}_n(z), {\cal H}_{n+1}(z), \ldots, {\cal H}_{n+m-1}(z)) = 
  {\cal W}\left(\{{\cal H}_{n+i}(z)\}_{i=0}^{m-1}\right), \quad m, n \ge 1.  \label{eq:gen-her-2}
\end{equation}
As special cases, we get $H_{1,n}(z) \propto \mathcal{H}_n(z)$ and 
$H_{2,n}(z) \propto \mathcal{W}( \mathcal{H}_n(z), \mathcal{H}_{n+1}(z))$.\par
%
%
\section{\boldmath COMPARISON BETWEEN $H_{1}$ AND ONE-STEP RATIONAL EXTENSIONS: ONE INFINITE SEQUENCE AND A SINGLET}
\setcounter{equation}{0}

In this section, we will relate the one-step rational extensions, reviewed in Eqs.~(\ref{eq:v1step}) to (\ref{eq:y1step}) of Sec.~IIA, together with their standard ladder operators $b^{\dagger}$, $b$, defined in Eq.~(\ref{eq:b}), with Hamiltonian $H_{1}$ of Eq.~(\ref{eq:h12p4}) for specific values of Painlev\'e IV parameters $\alpha$, $\beta$ (or, equivalently, for specific values of $\bar{\alpha}$ and $\sqrt{-d}$, as defined in (\ref{eq:para-andri})) and the ladder operators $a^+$, $a^-$, defined in Eq.~(\ref{eq:ap-am}). We will also show that the patterns of zero modes, the zero modes themselves, and the energy spectra coincide as well. To be able to carry out the comparison, we set $m_1=n$ ($n$ even) in Sec.~IIA.\par
%
%
{}For one-step rational extensions, the Hamiltonian reads
\begin{equation}
  H^{(2)} = - \frac{d^2}{dx^2} + x^2 - 2 \left[\frac{{\cal H}''_n}{{\cal H}_n} - \left(\frac{{\cal H}'_n}
  {{\cal H}_n}\right)^2 + 1 \right], 
\end{equation}
and its standard ladder operators take the form
\begin{equation}
\begin{split}
  & b^{\dagger} = \left(\frac{d}{dx} + W\right) \left(- \frac{d}{dx} + x\right) \left(- \frac{d}{dx} + W\right), \\ 
  & b = \left(\frac{d}{dx} + W\right) \left(\frac{d}{dx} + x\right) \left(- \frac{d}{dx} + W\right), \\
  & W(x) = - x - \frac{{\cal H}'_n}{{\cal H}_n}.
\end{split}  \label{eq:bpbw}
\end{equation}
As above-mentioned, for these ladder operators, the wavefunctions divide into one singlet state and one infinite sequence of equidistant levels. The energies of the singlet  and of the lowest state of the infinite sequence are given by $E^{(2)}_{-n-1} = -2n-1$ and $E^{(2)}_0 = 1$, with corresponding wavefunctions
\begin{equation}
  \psi^{(2)}_{-n-1}(x) \propto \frac{e^{-\frac{1}{2} x^2}}{{\cal H}_n}, \qquad \psi^{(2)}_0(x) \propto
  \frac{e^{-\frac{1}{2} x^2}}{{\cal H}_n} (- 2x {\cal H}_n - 2n {\cal H}_{n-1}), 
\end{equation}
respectively.\par
%
%
Let us now compare with Hamiltonian $H_{1}$ of Eq.~(\ref{eq:h12p4}), where we take for $g$ the following rational solution of Painlev\'e IV equation of type (\ref{eq:1wII}), connected with generalized Hermite polynomials,
\begin{equation}
  g(z) = w_{0.n}^{\rm II}(z) = \frac{d}{dz}\log\left(\frac{H_{1,n}}{H_{0,n}}\right) = \frac{d}{dz}
  \log \mathcal{H}_n = \frac{\mathcal{H}'_n}{\mathcal{H}_n},  \label{eq:g1}
\end{equation}
with corresponding parameters $\alpha_{0,n}^{(\rm II)}=n+1$, $\beta_{0,n}^{(\rm II)}=-2n^2$, coming from Table~II. Note that in the notations of Ref.~\cite{andri00}, the parameters now read $\bar{\alpha} = n$ and $\sqrt{-d} = n$ (see Eq.~(\ref{eq:para-andri})). For the choices $z=x$ and $c = 2\sqrt{-d} = 2n$, the three superpotentials, defined in (\ref{eq:w2}), become
\begin{equation}
  W_1 = W_3 = - x - \frac{{\cal H}'_n}{{\cal H}_n}, \qquad W_2 = x,  \label{eq:www1} 
\end{equation}
where we have taken Eq.~(\ref{eq:A3}) into account. Hence, the superpotentials $W_1$ and $W_3$ are equivalent to $W$, given in (\ref{eq:bpbw}), and the ladder operators $a^+$, $a^-$ of Eq.~(\ref{eq:ap-am}) coincide with $b^{\dagger}$, $b$ of Eq.~(\ref{eq:bpbw}). Furthermore, from Eqs.~(\ref{eq:haa}) and (\ref{eq:h1}), we also obtain
\begin{equation}
  H_1 = H^{(2)} + 2n + 1.  \label{eq:hh1}
\end{equation}
\par
%
%
Considering next the energies and wavefunctions of the zero modes, we get from Eqs.~(\ref{eq:energy0pm}) and (\ref{eq:energy123}), $E^{(0)}_0 = 0$, $E^{(0)}_+ = 2n+2$, and $E_1 = 0$, which are to be compared with $E^{(0)}_0 = E^{(2)}_{-n-1} + 2n + 1$, $E^{(0)}_+ = E^{(2)}_0 + 2n + 1$, and $E_1 = E^{(2)}_{-n-1} + 2n + 1$, in agreement with Eq.~(\ref{eq:hh1}). With superpotentials (\ref{eq:www1}), the corresponding wavefunctions $\psi^{(0)}_0(x)$, $\psi^{(0)}_+(x)$, and $\psi_1(x)$ become
\begin{align}
  & \psi^{(0)}_0(x) = \psi_1(x) \propto \psi^{(2)}_{-n-1}(x) \propto \frac{e^{-\frac{1}{2} x^2}}{{\cal H}_n}, \\
  & \psi^{(0)}_+(x) \propto \psi^{(2)}_0(x) \propto \frac{e^{-\frac{1}{2} x^2}}{{\cal H}_n} 
      (- 2x {\cal H}_n - 2n {\cal H}_{n-1}).  \label{eq:psi0+} 
\end{align}
In deriving Eq.~(\ref{eq:psi0+}), we employ (\ref{eq:A1}) to write
\begin{equation}
  W_2 - W_3 = 2x + \frac{{\cal H}'_n}{{\cal H}_n} = \frac{2x {\cal H}_n + 2n {\cal H}_{n-1}}{{\cal H}_n}.
\end{equation}
We conclude that there is a complete equivalence between the three zero modes (two of the annihilation operator and one of the creation operator) obtained in both approaches. In that of Ref.~\cite{andri00}, such a pattern of zero modes corresponds to case (d).\par
%
%
\section{COMPARISON BETWEEN $H_{1}$ AND ONE-STEP RATIONAL EXTENSIONS: THREE INFINITE SEQUENCES}
\setcounter{equation}{0}

Let us now consider the same one-step rational extensions with a different set of ladder operators $c^{\dagger}$, $c$, defined in (\ref{eq:c}). As explained in Sec.~IIA, these ladder operators are of third order only for $m_1=2$. In such a special case, the Hamiltonian and its ladder operators read
\begin{equation}
  H^{(2)}= -\frac{d^2}{dx^2}+x^2+\frac{8}{2x^2+1}-\frac{16}{(2x^2+1)^{2}}-2
\end{equation}
and
\begin{equation}
\begin{split}
  & c^{\dagger} = \left(\frac{d}{dx} + W\right) \left(- \frac{d}{dx} + \bar{W}^{(1)}\right) \left(- \frac{d}{dx} +
        \bar{W}^{(2)}\right), \\ 
  & c = \left(\frac{d}{dx} + \bar{W}^{(2)}\right) \left(\frac{d}{dx} + \bar{W}^{(1)}\right) \left(- \frac{d}{dx} +
         W\right), \\
  & W(x) = - x - \frac{4x}{2x^2+1}, \qquad \bar{W}^{(1)}(x) = x - \frac{1}{x}, \qquad \bar{W}^{(2)}(x) = x
         + \frac{1}{x} - \frac{4x}{2x^2+1},
\end{split}  \label{eq:cpcw}
\end{equation}
respectively. The wavefunctions then divide into three infinite sequences. The energies of their lowest states, which are zero modes of $c$, are given by $E^{(2)}_{-3} = -5$, $E^{(2)}_1 = 3$, and $E^{(2)}_2 = 5$, with corresponding wavefunctions
\begin{equation}
\begin{split}
  & \psi^{(2)}_{-3}(x) \propto \frac{e^{-\frac{1}{2}x^2}}{{\cal H}_2}, \qquad \psi^{(2)}_1(x) \propto 
       \frac{e^{-\frac{1}{2}x^2}}{{\cal H}_2} (-{\cal H}_2 H_2 - 4{\cal H}_1 H_1), \\ 
  & \psi^{(2)}_2(x) \propto \frac{e^{-\frac{1}{2}x^2}}{{\cal H}_2} (-{\cal H}_2 H_3 - 4{\cal H}_1 H_2). 
\end{split}  \label{eq:c-zm}
\end{equation}
\par
%
%
Let us now take for the rational solution of Painlev\'e IV equation, involved in definition (\ref{eq:h12p4}) of Hamiltonian $H_1$, one of those given in Eq.~(\ref{eq:2wII}) and connected with generalized Okamoto polynomials, namely
\begin{equation}
  g(z) = w^{(\rm II)}_{1,0}(z) = - \frac{2}{3}z + \frac{d}{dz} \log\left(\frac{Q_{2,0}}{Q_{1,0}}\right) =
  - \frac{2}{3}z + \frac{4z}{2z^2+3} = - \frac{2z(2z^2-3)}{3(2z^2+3)},
\end{equation}
with corresponding parameters $\alpha^{(\rm II)}_{1,0} = 2$, $\beta^{(\rm II)}_{1,0} = - \frac{2}{9}$, coming from Table~III, or $\bar{\alpha} = 1$, $\sqrt{-d} = \frac{1}{3}$ for the choice of Ref.~\cite{andri00}.\par
%
%
On selecting this time $c = - 2\sqrt{-d} = - \frac{2}{3}$, the three superpotentials (\ref{eq:w2}), written in terms of the variable $z$, are given by
\begin{equation}
  W_1(z) = \frac{1}{3}z + \frac{1}{z} - \frac{4z}{2z^2+3}, \qquad W_2(z) = \frac{1}{3}z - \frac{1}{z}, \qquad
  W_3(z) = - \frac{1}{3}z - \frac{4z}{2z^2+3}.
\end{equation}
After an additional change of variable $z = \sqrt{3}\, x$, they become
\begin{equation}
\begin{split}
  & W_1(z(x)) = \frac{1}{\sqrt{3}} \left(x + \frac{1}{x} - \frac{4x}{2x^2+1}\right), \qquad W_2(z(x)) = 
  \frac{1}{\sqrt{3}} \left(x - \frac{1}{x}\right), \\
  & W_3(z(x)) = \frac{1}{\sqrt{3}} \left(-x - \frac{4x}{2x^2+1}\right), 
\end{split}
\end{equation}
or
\begin{equation}
  W_1(z(x)) = \frac{1}{\sqrt{3}} \bar{W}^{(2)}(x), \qquad W_2(z(x)) = \frac{1}{\sqrt{3}} \bar{W}^{(1)}(x),
  \qquad W_3(z(x)) = \frac{1}{\sqrt{3}} W(x).
\end{equation}
The ladder operators $a^+$, $a^-$ of Eq.~(\ref{eq:ap-am}), written in terms of the variable $z = \sqrt{3}\, x$, therefore coincide with $c^{\dagger}/(3\sqrt{3})$, $c/(3\sqrt{3})$, obtained from (\ref{eq:cpcw}).\par
%
%
{}Furthermore, from Eq.~(\ref{eq:h1}) we now get
\begin{equation}
  H_1 = - \frac{d^2}{dz^2} + W_3^2(z) + \frac{dW_3(z)}{dz} = \frac{1}{3} \left(- \frac{d}{dx^2} + W^2(x)
  + W'(x)\right),
\end{equation}
so that comparison with Eq.~(\ref{eq:haa}) yields
\begin{equation}
  H_1 = \frac{1}{3} \bigl(H^{(2)} + 5\bigr).  \label{eq:hh2}
\end{equation}
On the other hand, for the energies (\ref{eq:energy0pm}) of the zero modes of $a^-$, we obtain $E^{(0)}_0 = 0 = \frac{1}{3}\bigl(E^{(2)}_{-3} + 5\bigr)$, $E^{(0)}_+ = \frac{8}{3} = \frac{1}{3}\bigl(E^{(2)}_{1} + 5\bigr)$, and $E^{(0)}_- = \frac{10}{3} = \frac{1}{3}\bigl(E^{(2)}_2 + 5\bigr)$, in agreement with those of $c$ and with Eq.~(\ref{eq:hh2}).\par
%
%
{}Finally, on considering the wavefunctions (\ref{eq:0pm}) of the zero modes, we note that the three exponentials are easily calculated and that the multiplying factors reduce to
\begin{equation}
  W_2(z(x)) - W_3(z(x)) = \frac{1}{2\sqrt{3}\, x{\cal H}_2} ({\cal H}_2 H_2 + 4{\cal H}_1 H_1),
\end{equation}
and
\begin{equation}
  - \frac{2}{3} + [W_2(z(x)) - W_3(z(x))] [W_1(z(x)) + W_2(z(x))] = \frac{2x}{3 {\cal H}_2^2} ({\cal H}_2 H_3 
  + 4{\cal H}_1 H_2),
\end{equation}
after some straightforward calculations using the explicit expressions of $H_1$, $H_2$, $H_3$, ${\cal H}_1$, and ${\cal H}_2$. This leads to
\begin{equation}
  \psi^{(0)}_0(z(x)) \propto \psi^{(2)}_{-3}(x), \qquad \psi^{(0)}_+(z(x)) \propto \psi^{(2)}_1(x), \qquad
  \psi^{(0)}_-(z(x)) \propto \psi^{(2)}_2(x),
\end{equation}
where the right-hand sides are given in Eq.~(\ref{eq:c-zm}). This completes the comparison between both approaches and the proof of their equivalence. In Ref.~\cite{andri00}, the pattern of zero modes observed here corresponds to case (a).\par
%
%
\section{\boldmath COMPARISON BETWEEN $H_1$ AND TWO-STEP RATIONAL EXTENSIONS: ONE INFINITE SEQUENCE AND A DOUBLET}
\setcounter{equation}{0}

In this section, we will relate the two-step rational extensions, considered in Sec.~IIB, together with their ladder operators $d^{\dagger}$, $d$, defined in (\ref{eq:dpd}), with Hamiltonian $H_1$ of Eq.~(\ref{eq:h12p4}) and its ladder operators $a^+$, $a^-$, defined in (\ref{eq:ap-am}). As above-mentioned, $d^{\dagger}$ and $d$ are third-order operators if we restrict ourselves to the case  where $m_2 = m_1+1$. To carry out the comparison between both approaches, let us reset $m_1=n$ and $m_2=n+1$ (with $n$ assumed even).\par
%
%
In these notations, the Hamiltonian for two-step rational extensions and its ladder operators read
\begin{equation}
  H^{(2)} = - \frac{d^2}{dx^2} + x^2 - 2 \left[\frac{g''_{2n}}{g_{2n}} - \left(\frac{g'_{2n}}{g_{2n}}\right)^2 +
  2 \right], \qquad g_{2n}(x) = {\cal W}({\cal H}_n, {\cal H}_{n+1}), \label{eq:h2step}
\end{equation}
and
\begin{equation}
\begin{split}
  & d^{\dagger} = \left(\frac{d}{dx} + W^{(2)}\right) \left(- \frac{d}{dx} + \hat{W}_1\right) 
       \left(- \frac{d}{dx} + \tilde{W}^{(2)}\right), \\ 
  & d = \left(\frac{d}{dx} + \tilde{W}^{(2)}\right) \left(\frac{d}{dx} + \hat{W}_1\right) \left(- \frac{d}{dx} +
       W^{(2)}\right), \\
  & W^{(2)}(x) = - x + \frac{{\cal H}'_n}{{\cal H}_n} - \frac{g'_{2n}}{g_{2n}}, \\
  & \hat{W}_1(x) = x + \frac{{\cal H}'_n}{{\cal H}_n} - \frac{{\cal H}'_{n+1}}{{\cal H}_{n+1}}, \\
  & \tilde{W}^{(2)}(x) = - x + \frac{{\cal H}'_{n+1}}{{\cal H}_{n+1}} - \frac{g'_{2n}}{g_{2n}}, 
\end{split}  \label{eq:super2step1}
\end{equation}
respectively. The annihilation operator $d$ has two zero modes, which are the lowest states of the doublet and of the infinite sequence, while the creation operator $d^{\dagger}$ has one zero mode, which is the highest state of the doublet. The corresponding energies and wavefunctions are given by
\begin{equation}
  E^{(2)}_{-n-2} = - 2n - 3, \qquad E^{(2)}_0 = 1, \qquad E^{(2)}_{-n-1} = - 2n - 1, \label{eq:modes2step}
\end{equation}
and
\begin{equation}
\begin{split}
  & \psi^{(2)}_{-n-2}(x) \propto \frac{e^{-\frac{1}{2}x^2}}{g_{2n}} {\cal H}_n, \\
  & \psi^{(2)}_0(x) \propto \frac{e^{-\frac{1}{2}x^2}}{g_{2n}} \{{\cal H}_n {\cal H}_{n+1} H_1 + 2[n(n+2)
       {\cal H}_{n-1} {\cal H}_{n+1} - (n+1)^2 {\cal H}_n^2]\}, \\
  & \psi^{(2)}_{-n-1}(x) \propto \frac{e^{-\frac{1}{2}x^2}}{g_{2n}} {\cal H}_{n+1},  \label{eq:psi2step}
\end{split}
\end{equation}
respectively.\par
%
%
{}For the rational solution of Painlev\'e IV equation entering definition (\ref{eq:h12p4}) of $H_1$, let us choose one of the functions (\ref{eq:1wII}) connected with generalized Hermite polynomials,
\begin{equation}
  g(z) = w^{(\rm II)}_{1,n}(z) = \frac{d}{dz} \log\left(\frac{H_{2,n}}{H_{1,n}}\right) = \frac{d}{dz} \log\left(
  \frac{g_{2n}}{{\cal H}_n}\right) = \frac{g'_{2n}}{g_{2n}} - \frac{{\cal H}'_n}{{\cal H}_n}, \label{eq:g2step}
\end{equation}
where, in the second step, use is made of Eq.~(\ref{eq:gen-her-2}). From Table II, we obtain for the corresponding parameters $\alpha^{(\rm II)}_{1,n} = n+3$ and $\beta^{(\rm II)}_{1,n} = -2n^2$. In the notations of Ref.~\cite{andri00}, these become $\bar{\alpha} = n+2$ and $\sqrt{-d} = n$. For the choices $z=x$ and $c = 2\sqrt{-d} = 2n$, the three superpotentials of Eq.~(\ref{eq:w2}) read
\begin{equation}
  W_1(x) = - \frac{g}{2} + \frac{g'}{2g} - \frac{n}{g}, \qquad W_2(x) = - \frac{g}{2} - \frac{g'}{2g} + 
  \frac{n}{g}, \qquad W_3(x) = - x - g,  \label{eq:super2step2}
\end{equation}
with $g$ given in (\ref{eq:g2step}).\par
%
%
Comparison  between the two sets of superpotentials given in (\ref{eq:super2step1}) and (\ref{eq:super2step2}) immediately leads to the relations
\begin{equation}
  W_3(x) = W^{(2)}(x) = - x - g, \qquad W_1(x) + W_2(x) = \hat{W}_1(x) + \tilde{W}^{(2)}(x) = - g.
\end{equation}
To show the equivalence of $(W_1, W_2, W_3)$ and $(\tilde{W}^{(2)}, \hat{W}_1, W^{(2)})$, it remains to prove that $W_1 = \tilde{W}^{(2)}$ or, equivalently, $2g W_1 - 2g \tilde{W}^{(2)} = 0$. Inserting Eq.~(\ref{eq:g2step}) and its derivative in the left-hand side of this relation transforms the latter into
\begin{equation}
  2g W_1 - 2g \tilde{W}^{(2)} = \frac{g''_{2n}}{g_{2n}} + 2x\frac{g'_{2n}}{g_{2n}} - 
  \frac{{\cal H}''_n}{{\cal H}_n} - 2x\frac{{\cal H}'_n}{{\cal H}_n} - 2 \frac{{\cal H}'_{n+1}}{{\cal H}_{n+1}}
  \frac{g'_{2n}}{g_{2n}} + 2 \frac{{\cal H}'_n}{{\cal H}_n} \frac{{\cal H}'_{n+1}}{{\cal H}_{n+1}} - 2n.
  \label{eq:relation}
\end{equation}
On using the recursion and differential equations satisfied by ${\cal H}_n$ and $g_{2n}$ and listed in Appendix A, it is then straightforward to prove the vanishing of Eq.~(\ref{eq:relation}) right-hand side. We therefore conclude that the ladder operators $a^+$, $a^-$ of Eq.~(\ref{eq:ap-am}) coincide with $d^{\dagger}$, $d$ of Eq.~(\ref{eq:super2step1}).\par
%
%
{}Furthermore, combining Eqs.~(\ref{eq:h12p4}) and (\ref{eq:g2step}) yields
\begin{equation}
  H_1 = - \frac{d^2}{dx^2} + x^2 - \frac{g''_{2n}}{g_{2n}} + 2 \left(\frac{g'_{2n}}{g_{2n}}\right)^2
  + 2x \frac{g'_{2n}}{g_{2n}} + \frac{{\cal H}''_n}{{\cal H}_n} - 2x \frac{{\cal H}'_n}{{\cal H}_n} - 2
  \frac{g'_{2n}}{g_{2n}} \frac{{\cal H}'_n}{{\cal H}_n} - 1,
\end{equation}
from which we get
\begin{equation}
  H_1 = H^{(2)} + 2n + 3  \label{eq:hh2step}
\end{equation}
after using Eq.~(\ref{eq:h2step}) and the relations given in Appendix A again. On the other hand, from Eqs.~(\ref{eq:energy0pm}) and (\ref{eq:energy123}), we obtain $E^{(0)}_0 = 0 = E^{(2)}_{-n-2} + 2n + 3$, $E^{(0)}_+ = 2n + 4 = E^{(2)}_0 + 2n + 3$, $E_1 = 2 = E^{(2)}_{-n-1} + 2n + 3$, in agreement with Eqs.~(\ref{eq:modes2step}) and (\ref{eq:hh2step}).\par
%
%
It remains to check that the wavefunctions $\psi^{(0)}_0(x)$, $\psi^{(0)}_+(x)$, and $\psi_1(x)$, defined in Eqs.~(\ref{eq:0pm}) and (\ref{eq:123}), reduce to those previously obtained in (\ref{eq:psi2step}). The exponentials contained in (\ref{eq:0pm}) and (\ref{eq:123}) are easily calculated. This directly yields the relations
\begin{equation}
  \psi^{(0)}_0(x) \propto \psi^{(2)}_{-n-2}(x), \qquad \psi_1(x) \propto \psi^{(2)}_{-n-1}(x).
\end{equation}
Furthermore, $\exp\left(-\int^x W_2(x') dx'\right) = e^{-\frac{1}{2}x^2} {\cal H}_{n+1}/{\cal H}_n$. On the other hand, we can write
\begin{align}
  W_2 - W_3 &= \hat{W}_1 - W^{(2)} = 2x + \frac{g'_{2n}}{g_{2n}} - \frac{{\cal H}'_{n+1}}{{\cal H}_{n+1}}
      = 2 \frac{{\cal H}_n}{{\cal H}_{n+1} g_{2n}} [{\cal H}_{n+1}^2 - (n+1)g_{2n}] \nonumber \\
  & = 2 \frac{{\cal H}_n}{{\cal H}_{n+1} g_{2n}} \{2x {\cal H}_n {\cal H}_{n+1} + 2[n(n+2) {\cal H}_{n-1}
      {\cal H}_{n+1} - (n+1)^2 {\cal H}_n^2]\}
\end{align}
by using the relations of Appendix A. Hence,
\begin{equation}
  \psi^{(0)}_+(x) \propto \psi^{(2)}_0(x),
\end{equation}
which completes the proof of equivalence of both approaches, that of Ref.~\cite{andri00} corresponding this time to case (e).\par
%
%
\section{CONCLUSION}

In this paper, we discussed the connection between one- and two-step rational extensions $H^{(2)}$ of the harmonic oscillator and quantum systems $H_1$ involving rational solutions of the fourth Painlev\'e equation related to generalized Hermite and Okamoto polynomials. Since the wavefunctions of $k$-step rational extensions involve Hermite EOP of type III, the present study points out a nontrivial link among various special functions and orthogonal polynomials. \par
%
%
{}For three different cases exhibiting three different patterns of zero modes, the connection was established at the level of the Hamiltonians, supercharges, ladder operators, zero modes, and corresponding spectra. \par
%
%
In recent years, $k$-step extensions of the singular harmonic oscillator have been studied \cite{oda11, oda13, ull13, gomez12, cq11a, cq11b, grandati12, grandati13}, as well as some of their superintegrable generalizations \cite{marquette13a,marquette14}. On the other hand, some 1D quantum Hamiltonians \cite{carb04} and 2D superintegrable ones \cite{marquette11} involving the fifth Painlev\'e transcendent have been considered. Establishing the connection between both approaches would be a very interesting topic for future investigation, although this is a much harder problem than that solved in the present paper, because the special solutions of the fifth Painlev\'e equation \cite{clar05} and their asymptotic are more complicated. \par
%
%
\section*{ACKNOWLEDGMENTS}

The research of I.M.\ was supported by the Australian Research Council through Discovery Early Career Researcher Award DE130101067. I.M.\ thanks the Universit\'e Libre de Bruxelles for its hospitality. \par
%
%
\section*{\boldmath APPENDIX A: SOME IDENTITIES SATISFIED BY ${\cal H}_n$ AND $g_{2n}$}

\renewcommand{\theequation}{A.\arabic{equation}}
\setcounter{equation}{0}

We provide here a list of identities \cite{marquette13b} used to demonstrate various elements of the equivalence:
\begin{equation}
  \mathcal{H}'_{n}=2n\mathcal{H}_{n-1}, \label{eq:A1}
\end{equation}
\begin{equation}
  \mathcal{H}'_{n}+2x\mathcal{H}_{n}=\mathcal{H}_{n+1},  \label{eq:A2}
\end{equation}
\begin{equation}
  \mathcal{H}''_{n}+2x\mathcal{H}'_{n}-2n\mathcal{H}_{n}=0,   \label{eq:A3}
\end{equation}
\begin{equation}
  g_{2n}'+2xg_{2n}=2\mathcal{H}_{n}\mathcal{H}_{n+1},   \label{eq:A4}
\end{equation}
\begin{equation}
  g_{2n}''+2xg_{2n}'= 4 {\cal H}'_n {\cal H}_{n+1}.    \label{eq:A5}
\end{equation}
\par
%
%

\end{document}